\documentclass[12pt,preprint]{aastex}

\usepackage{amssymb}
\usepackage{amsmath}

\shorttitle{Very first stars and z\~6 QSOs} 
\shortauthors{Trenti \& Stiavelli}

\begin{document}


\title{Distribution of the very first PopIII stars and their relation
to bright $z\approx 6$ quasars}


\author{M. Trenti} 
\affil{Space Telescope Science Institute, 3700 San Martin Drive Baltimore MD 21218 USA}
\and 
\author{M. Stiavelli}
\affil{Space Telescope Science Institute, 3700
San Martin Drive Baltimore MD 21218 USA; \\ Department of Physics and
Astronomy, Johns Hopkins University, Baltimore, MD 21218 USA }
\email{trenti@stsci.edu; mstiavel@stsci.edu}


\begin{abstract}

We discuss the link between dark matter halos hosting the \emph{first}
PopIII stars and the rare, massive, halos that are generally
considered to host bright quasars at high redshift ($z \approx
6$). The main question that we intend to answer is whether the
super-massive black holes powering these QSOs grew out from the seeds
planted by the \emph{first} intermediate massive black holes created
in the universe. This question involves a dynamical range of $10^{13}$
in mass and we address it by combining N-body simulations of structure
formation to identify the most massive halos at $z \approx 6$ with a
Monte Carlo method based on linear theory to obtain the location and
formation times of the first light halos within the whole simulation
box. We show that the descendants of the first $\approx 10^6 M_{\sun}$
virialized halos do not, on average, end up in the most massive halos
at $z \approx 6$, but rather live in a large variety of
environments. The oldest PopIII progenitors of the most massive halos
at $z \approx 6$, form instead from density peaks that are on average
one and a half standard deviations more common than the first PopIII
star formed in the volume occupied by one bright high-z QSO. The
intermediate mass black hole seeds planted by the very first PopIII
stars at $z\gtrsim 40$ can easily grow to masses $m_{BH}>10^{9.5}
M_{\sun}$ by $z=6$ assuming Eddington accretion with radiative
efficiency $\epsilon \lesssim 0.1$. Quenching of the black hole
accretion is therefore crucial to avoid an overabundance of
supermassive black holes at lower redshift. This can be obtained if
the mass accretion is limited to a fraction $\eta \approx 6 \cdot
10^{-3}$ of the total baryon mass of the halo hosting the black
hole. The resulting high end slope of the black hole mass function at
$z=6$ is $\alpha \approx -3.7$, a value within the $1\sigma$ error bar
for the bright end slope of the observed quasar luminosity function at
$z=6$.
 
\end{abstract}

\keywords{cosmology: theory - galaxies: high-redshift - early universe
- methods: N-body simulations}

\section{Introduction}

Bright quasars at $z \approx 6$ are very luminous and rare objects that
can can be detected out to huge cosmological distances in very large
area surveys like the Sloan Digital Sky Survey \citep{fan04}. Their
estimated space density is $\approx 2.2 \cdot 10^{-9} (Mpc/h)^{-3}$
\citep{fan04}, that is about one object per about $200$ deg$^2$ of sky,
assuming a depth of $\Delta z = 1$ centered at $z=6.1$ under the third
year WMAP cosmology \citep{WMAP3}. Their luminosity is thought to be
due to accretion onto a super-massive black hole \citep[e.g.,
see][]{hop05}.
A common expectation is that the luminous high-z quasars sit at the
center of the biggest proto-clusters at that time. Some observational
evidence of over-densities of galaxies in two deep HST-ACS fields
containing a bright z=6 quasar has been claimed \citep{sti05,zhe06},
but it is unclear whether this is true in general. In fact an ACS
image only probes a long and narrow field of view of about $6 \times 6
\times 320 (Mpc/h)^3$ in the redshift range $[5.6:6.6]$, so a
significant number of detections may come from galaxies unrelated to
the environment of the host halo of the bright quasar.

Numerical simulations to address the formation of bright quasars are
extremely challenging given their low number density. A huge
simulation cube with edge of $\gtrsim 700~Mpc/h$ is required just to
expect, on average, one such object in the simulation box. A major
computational investment, like the Millennium run \citep{MR05}, is
required to resolve at high redshift ($z \gtrsim 20$) virialized halos
on this volume and to follow their merging history down to $z \approx
6$. Even assuming that the simulation volume is big enough that there
is the expectation to find halos hosting bright quasars, how can these
halos be identified? In principle two, non mutually exclusive,
alternatives appear plausible: either the super-massive black holes
are hosted in the most massive halos with the corresponding number
density of SDSS quasars or these black holes have grown from the first
PopIII Intermediate Mass Black Hole seeds, therefore representing the
descendant of the rarest density peaks that hosted first stars.

The first scenario implies that the $m_{BH}-\sigma$ relation
\citep{fer00,geb00} is already in place at high redshift
\citep{vol03,hop05,dimatteo05}. In that case multigrid simulations can
be carried out to follow in detail the growth of the supermassive
black hole (e.g., see \citealt{li06}). In the second scenario the
quasars progenitors would be traced back to the first PopIII stars
created in the universe within $\approx 10^6 M_{\sun}$ mass halos
virialized at $z \approx 50$ \citep{bro04,abel02}. These PopIII stars
are very massive $M>100M_{\sun}$, so after a short life of a few
million years explode and may leave intermediate mass black holes,
plausible seeds for the super-massive black holes observed at lower
redshift. Of course the two scenarios can be consistent with each
other if the first perturbations to collapse are also the most massive
at $z=6$. This seems to be implied, e.g. in \citet{MR05}, where the
bright quasar candidate in the simulation is traced back to one of the
18 collapsed halos at $z=16.7$.

In this paper we explore the link between the first PopIII halos
collapsed in a simulation box and the most massive halos at lower
redshifts to gain insight on the scenarios of bright quasar
formation. This is a numerically challenging problem as the dynamical
range of masses involved is very large: a simulation volume of $5
\cdot 10^{8} (Mpc/h)^{3}$ has a mass of about $3.3 \cdot 10^{19}
M_{\sun}/h$, that is more than $10^{13}$ times the mass of a PopIII
dark matter halo. We have adopted an original approach to the problem,
broadly inspired by the tree method by \citet{col94}. We first
simulate at relatively low resolution the evolution of a simulation
volume down to $z=0$. Then, starting from the density fluctuations
field in the initial conditions of the numerical simulation, we
compute analytically the redshift distribution of the oldest PopIII
halo collapsed within each single grid cell. The information is then
used as input for a Monte Carlo code to sample for each particle of
the simulation the collapse redshift of the first PopIII progenitor
dark matter halo. The formation time of the oldest PopIII remnant
within the most massive halos identified at $z \approx 6$ is finally
compared with that of the oldest PopIII star sampled over the whole
simulation volume and the implications for the growth of supermassive
black holes are discussed.

Our approach is tuned to investigate the formation and the subsequent
remnant distribution of the first, rare density peaks that hosted
PopIII stars at $z \gtrsim 30$.  With this respect our study has a
similar goal to \citet{ree05}, with the important difference that we
search for the first PopIII star in the complete simulation box and
not by means of progressive refinements around substructures that
probe only a small fraction of the total box volume. As our method is
tuned at finding very rare fluctuations, it is not easily applied to
the significantly more common $3 \sigma$ peaks with mass
$\approx 10^{6}M_{\sun}$ that collapse at $z\approx 20$ and that might
constitute the majority of PopIII stars, if these are terminated by
chemical feedback at $z \lesssim 20$ \citep[e.g.  see,][]{gre06} and
not by photo-dissociation of molecular hydrogen at $z \gtrsim 25$
\citep{hai00}.

This paper is organized as follows. In Sec.~\ref{sec:num} we present
the details of the numerical simulations carried out. In
Sec.~\ref{sec:first_halos} we analyze the numerical results focusing
on the merging history of the first PopIII halos formed in the
simulation box. In Sec.\ref{sec:end_of_fs} we review when the first
stars epoch end, while in Sec.~\ref{sec:BHgrow} we discuss the
implications of the PopIII distribution that we find for the build-up
of supermassive black hole population at $z \lesssim 6$. We conclude
in Sec.~\ref{sec:conc}.

\section{Numerical Methods} \label{sec:num}

\subsection{N-body simulations}

The numerical simulations presented in this paper have been carried
out using the public version of the PM-Tree code Gadget-2
\citep{spr05}. Our standard choice is to adopt a cosmology based on
the third year WMAP data \citep{WMAP3}: $\Omega_{\Lambda}=0.74$,
$\Omega_{m}=0.26$, $H_0=70~km/s/Mpc$, where $\Omega_m$ is the total
matter density in units of the critical density ($\rho_{c}= 3H_0^2/(8
\pi G)$) with $H_0$ being the Hubble constant (parameterized as $H_0 =
100 h~ km/s/Mpc$) and $G$ the Newton's gravitational constant
\citep{peebles}. $\Omega_{\Lambda}$ is the dark energy density.  As
for $\sigma_8$, the root mean squared mass fluctuation in a sphere of
radius $8Mpc/h$ extrapolated at $z=0$ using linear theory, we consider
both $\sigma_8=0.9$ and $\sigma_8=0.75$, focusing in particular on the
higher value that provides a better match to the observed clustering
properties of galaxies \citep{evr07}.

The initial conditions have been generated using a code based on the
Grafic algorithm \citep{bert01}. An initial uniform lattice is
perturbed using a discrete realization of a Gaussian random field
sampled in real space and then convolved in Fourier space with a
$\Lambda CDM$ transfer function computed using the fit by \citet{eis99}
and assuming a scale invariant long-wave spectral index ($n=1$). The
initial density field is saved for later reprocessing through the
first light Monte Carlo code (see Sec.~\ref{sec:MC}). The particles
velocities and displacements are then evolved to the desired starting
redshift ($z_{start}=65.67$, i.e. $a_{start}=0.015$) using the
Zel'dovich approximation and the evolution is followed using Gadget-2
\citep{spr05}. Dark matter halos are identified in the simulations
snapshots using the HOP halo finder \citep{eis98}.

To find the optimal trade-off between mass resolution and box size,
both critical parameters to establish a connection between PopIII
halos and the most massive halos identified at $z=6$, we resort to
simulations (see Tab.~\ref{tab:sim}) with three different box sizes,
all simulated with $N=512^3$ particles:
\begin{itemize}

\item[(i)] A ``large'' box size of edge $720~Mpc/h$ that is large
enough to contain on average about one bright high-z quasar. The mass
resolution is $3.7 \cdot 10^{12} M_{\sun}/h$ (corresponding to a halo
of 20 particles).

\item[(ii)] A ``medium'' box size of edge $512~Mpc/h$ that represents
a compromise between a slightly higher mass resolution than $(i)$ and
a still reasonably large simulation volume.

\item[(iii)] A ``small'' box size of edge $60~Mpc/h$. While this box
  size is too small to host a bright $z \approx 6$ quasar, its volume
  is still larger than that of deep surveys like the UDF
  \citep{beck06}, that spans a volume about $20$ times smaller than
  this box in the redshift interval $z \in [5.6:6.6]$ (the typical
  redshift uncertainty for $i$-dropouts). Halos down to about $3 \cdot
  10^{9} M_{\sun}/h$ can be identified in this box. The analysis of
  the results from this simulation will show the fundamental role
  played by the large volume employed for simulations (i) and (ii). 

\end{itemize}

\subsection{Monte Carlo code for first light sources}\label{sec:MC}

Given the initial density fluctuations field on the simulation grid,
where a cell has a mass of order $10^{10}- 10^{11} M_{\sun}/h$, our
goal is to estimate the redshift of the first virialized perturbation
within each cell at the mass scale of early PopIII dark matter halos
(i.e. $\lesssim 10^6 M_{\sun}$, see e.g. \citealt{bro04}). For this we
resort to an analytical treatment based on a linear approximation for
structure formation.

The initial conditions for a N-body simulation in a box of size $L$
with $N$ particles and a single particle mass $m_p$ define a Gaussian
random field $\delta {\rho}$ for the N cells (associated to the
location of the N particles) of the simulation grid. This density
field is usually generated by convolving white noise with the transfer
function associated to the power spectrum of the density perturbations
(e.g., see \citealt{bert01}) and is used to obtain the initial
velocity and positions displacements for the particles (e.g. see
Eq.~5.115 in \citealt{peebles}). The density fluctuation in each cell
has a contribution from different uncorrelated frequencies in the
power spectrum. When the initial conditions for an N-body simulation
are generated, the power spectrum has an upper cutoff around the
Nyquist frequency for the grid used (i.e. around the frequency
associated to the average inter-particle distance) and a lower cutoff
at the frequency associated to the box size (if periodic boundary
conditions are enforced). A higher resolution version of the initial
density field can be obtained by simply increasing the grid size and
adding the density perturbations associated to the power spectrum
between the old and the new cutoff frequencies.

In linear approximation one can use the field $\delta \rho$ to obtain
the redshift of virialization of a structure of mass $M_h>m_p$ at a
given position $\vec{x}$ in the grid. To do this one averages the
field $\delta \rho$ using a spherical window centered at $\vec{x}$
with a radius such that the window encloses a mass $M_h$ and computes
assuming linear growth the redshift at which the average density
within the window reaches $\delta \rho = 1.69$ (in units where the
average density of the box is 1). In fact, for a spherical collapse
model, when $\delta \rho = 1.69$ in linear theory, then the halo has
reached virial equilibrium under the full non-linear dynamics. This
concept is at the base of the various proposed methods for computing
analytically the mass function of dark matter halos (e.g., see
\citealt{PS,bond,she99}).

We apply this idea to estimate the formation rate and the location in
the simulation volume of dark matter halos at a mass scale below the
single particle mass used in the simulation. A straightforward
implementation consists in generating first the density field
associated to the N-body simulation, and then to refine at higher
resolution the field by means of a constrained realization of the
initial conditions used in the N-body run (e.g. see
\citealt{bert01}). This provides exact and complete information on the
whole density field, but the price to pay is the execution of very
large Fast Fourier Transforms on the refined grid. If the goal is to
compute density fluctuations down to a mass of $\approx 10^6
M_{\sun}/h$ over a box of edge $720 Mpc/h$, a grid of $29184^3$ is
needed, which would require about $100$ TB of RAM, that is well beyond
the current memory capabilities of the largest supercomputers.
 
A shortcut is however available, if one trades information for
numerical complexity. Given a realized numerical simulation, we are in
fact not interested in getting a detailed picture of the dynamics at
sub-grid resolution, but only in identifying for each grid point the
redshift of virialization of its first progenitor at a given sub-grid
mass scale. For example, given a simulation with single particle mass
of $10^{10} M_{\sun}/h$ our aim is to quantify the redshift of
virialization of the \emph{first} dark matter halo of mass $10^6
M_{\sun}/h$ within the volume associated to the $10^{10} M_{\sun}/h$
particle. In that case, if we were to have the full sub-grid
information we would search for the maximum realized value of the
density within the $10^4$ sub-grid cells of mass $10^6 M_{\sun}/h$
that constitute our $10^{10} M_{\sun}/h$ single particle cell. As the
density fluctuation field is a Gaussian random field, the density in
sub-grid cells will be a Gaussian centered at the density of the
parent cell and with variance given by integration of the power
spectrum of density fluctuations truncated between the Niquist
frequency of the parent cell and that of the sub-cells.

Therefore, for a single cell of mass $m_p$, the redshift of collapse
of a sub-grid progenitor at a mass scale $m_{fs}$ can be obtained
simply by sampling from the probability distribution of the maximum of
the sub-grid fluctuations of the $k=m_p/m_{fs}$ sub-cells of mass
$m_{fs}$ that are within a cell of mass $m_p$. 
The probability distribution for the maximum of these
fluctuations is available in analytic form when the field is Gaussian,
as in the case considered here. 
In fact, given a probability distribution $p(x)$, with partition function
$P(x)$:
\begin{equation}
P(x)=\int_{- \infty}^x p(a) da, 
\end{equation}
the probability distribution $q(m,k)$ for the maximum $m$ of $k$
random numbers extracted from $p(x)$ ($m=Max(x_1,...,x_k)$) is the
derivative of the partition function for $m$, that in turns is simply
the $k$-th power of the partition function for $x_i$,
i.e. $P(x)$. Therefore we have:
\begin{equation} \label{eq:maxpdf}
q(m,k) = k \cdot p(m) \cdot P(m)^{k-1}. 
\end{equation}
Eq.~(\ref{eq:maxpdf}) has a simple interpretation: the probability
that the maximum of $k$ random numbers lies in the interval
$[m,m+\delta m]$ is given by the probability of sampling one of the $k$
numbers exactly in that interval and all the other numbers below $m$.

With the aid of Eq.~(\ref{eq:maxpdf}) we can sample the distribution
of the maximum of the additional sub-grid density fluctuations that
need to be considered in order to probe the mass scale of PopIII
halos. The variance $\sigma_{fs}$ to be used in $p(x)$ may be computed
from the power spectrum of the density fluctuations by considering an
upper cut-off at the wavelength of one cell size in the initial
conditions. Or, equivalently, if the complete power spectrum of
density fluctuations has variance $\sigma(M_{grid\_cell})$ at the mass
scale of one grid cell and variance $\sigma(M_{PopIII\_halo})$ at the
mass scale of a halo hosting a first star, we set $\sigma_{fs}$ such
that:
\begin{equation} \label{eq:sigma}
\sigma_{fs}^2 = \sigma(M_{PopIII\_halo})^2 - \sigma(M_{grid\_cell})^2.
\end{equation}

Therefore our recipe for estimating the age of the earliest progenitor
formed in each cell is the following:
\begin{itemize}
\item[(i)] Starting from the initial density fluctuation field on the
grid used to initialize the N-body run compute the mass refinement
factor $k$ to go from the mass of a single particle (i.e. the mass
within one grid cell) to that of a PopIII star halo
($k=M_{grid\_cell}/M_{PopIII\_halo}$).
\item[(ii)] Given the power spectrum of the density fluctuations,
$M_{grid\_cell}$ and $M_{PopIII\_halo}$ compute $\sigma_{fs}$. 
\item[(iii)] Extract one random number $r$ from $q(m,k)$ (see
Eq.~\ref{eq:maxpdf}) where $p(m)$ is a Gaussian distribution with zero
mean and variance $\sigma_{fs}$.
\item[(iv)] Sum $r$ to the value of the density field in the cell to
obtain $(\delta \rho_{fs} / \rho)_{max}$ in the cell. From the value
of $(\delta \rho_{fs}/ \rho)_{max}$ it is then straightforward to
compute the non-linear redshift for that perturbation, i.e. the
redshift $z_{nl}$ when the linear density contrast reaches a value
$(\delta \rho_{fs} / \rho)_{max} (z_{nl}) = 1.69$.
\end{itemize}

The particles of the simulation now carry the additional information
of the redshift at which their \emph{first} PopIII star dark matter
halo progenitor has collapsed in linear theory (a proxy for the
redshift of actual virialization). Once halos have been identified in
simulation snapshots, the redshift of the earliest PopIII progenitor
within the halo is easily obtained. It is similarly easy to identify
in a snapshot what is the environment in which the particles with the
oldest progenitors live. This procedure is robust with respect to
variations of the simulation resolution, as long as the focus is on
rare density peaks, with an average occupation number per simulation
cell (i.e. particle) much smaller than unity. Numerical tests are
presented in Appendix~\ref{sec:app}.

This method has two main advantages: 
\begin{enumerate}
\item[(i)] It allows to use relatively inexpensive ``low resolution''
simulations to identify the largest objects at low redshift ($z
\lesssim 6$). In fact if we are interested in identifying the most
massive halos at $z \approx 6$ as host halos for quasar candidates a
mass resolution of $\approx 10^{11} M_{\sun}/h$ is sufficient (e.g. in
\citealt{MR05} the mass of the largest halo at $z=6.2$ is $3.9 \cdot
10^{12} M_{\sun}/h$ for a simulation volume of $(500 Mpc/h)^3$).

\item[(ii)] For a given numerical simulation, several Monte Carlo
realization can be generated to gather robust statistical constraints
on the properties of dark matter halos hosting first light sources as
well as the spatial distributions of the first halo remnants in halos
at lower redshift.

\end{enumerate}

However our method has the drawback that it cannot be easily extended
to the investigation of the detailed merging history at the sub-grid
level, as only the virialization time of the earliest progenitor of
each particle at a given mass scale is provided. In addition, the
identification of the first virialized PopIII halos is expressed in
terms of the halos with the highest $z_{nl}$. We are therefore
neglecting the non-linear evolution and the environmental dependences
on the dynamics of the dark matter collapse, such as tidal forces,
therefore missing the precise redshift at which a PopIII halo
virializes. These are limitations that we need to accept as the non
linear evolution could be followed over the whole box only at the
price of running a simulation prohibitively intensive in cpu and
memory resources, with at least $10^3$ time more particles than in the
Millenium Run \citep{MR05}. This appears unfeasible for the time
being, even considering next generation dedicated supercomputers, like
the GrapeDR \citep{mak05}.

\section{The fate of the first PopIII halos} \label{sec:first_halos}

\subsection{Analytical considerations}\label{sec:est}

The general picture for the connection between first halos and the
most massive halos at $z \approx 6$ can be obtained using analytical
considerations, that will be later confirmed in Sec.~\ref{sec:MCres}
by the results of our numerical investigation. 

\begin{figure} 
  \plotone{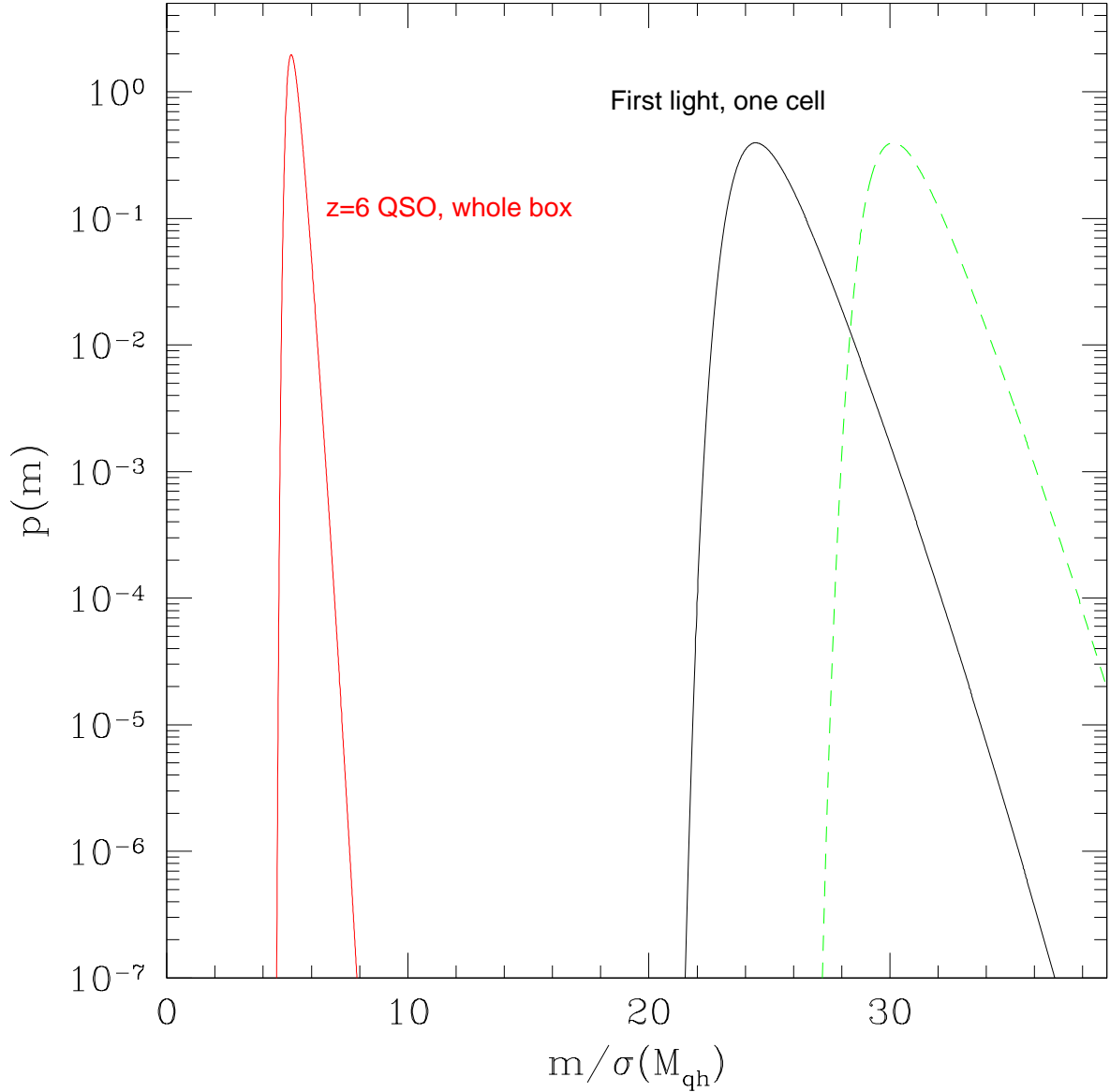} \caption{Probability distribution functions
  $p(m)$ for the maximum $m$ of $k$ random Gaussian fluctuations
  representative of mass scale for halos hosting $z=6$ QSO candidates
  (red curve) and PopIII stars (black curve). To compute $p(m)$ for
  QSO hosting halos we have assumed a box of $720Mpc/h$ and a mass
  scale of $M_{qh} = 4.3~10^{12} M_{\sun}/h$ which implies
  $k=180^3$. This curve represents the probability distribution for a
  sigma peak that at $z \approx 6$ leads to one of the most massive
  halos in the simulation volume. The curve associated to first light
  perturbations (solid black, with $M_{fs}=10^{6} M_{\sun}/h$) is
  derived using
  $\sigma_{fs}=\sqrt{\sigma_{M_{fs}}^2-\sigma_{M_{qh}}^2}$ and
  $k=160^3$: it represents the probability distribution for the
  maximum of the sub-grid scale fluctuations at the $M_{fs}$ scale
  within one cell of the $180^3$ volume. The dashed green line
  represents the probability distribution of the density fluctuation
  associated to the first PopIII progenitor of a $5.7 \sigma(M_{qh})$
  peak. $m$ is given in units of $\sigma_{M_{qh}}$.}\label{fig:pdf}
\end{figure}

Following the choice for our large box simulation, we consider a
volume of $(720 Mpc/h)^3$ of mass $M_{Box}$, large enough to host a
bright $z \approx 6$ quasar. We estimate from the Press-Schechter
formalism (see also the masses of the $z=6$ halos in our ``large'' box
simulation in Sec.~\ref{sec:MCres}) that the most massive halo at $z
\approx 6$ has a mass (that we call $M_{qh}$) of about $10^{12} -
10^{13} M_{\sun}/h$ (see also \citealt{MR05}). Since the most massive
halo is the first at its mass scale to be formed, through the use of
Eq.~\ref{eq:maxpdf} we can obtain the distribution of its initial
density fluctuation (see Fig.~\ref{fig:pdf}). If we assume $M_{qh} =
M_{Box}/180^3 = 4.3 \cdot 10^{12} M_{\sun}/h$ (in agreement with
\citealt{MR05}), this halo is expected to have originated from a
density fluctuation in the range $[5:5.7] \sigma(M_{qh})$ at $90 \%$
of confidence level. We now consider the volume initially occupied by
the mass $M_{qh}$ and we compute from the primordial power spectrum
the variance $\sigma_{fs}$ of density perturbations at mass scale of a
PopIII halo ($M_{fs} = 10^6 M_{\sun}/h = 1/160^3 M_{qh}$) considering
only contributions from wavelengths at a scale below the volume
enclosed by $M_{qh}$ (see Eq.~\ref{eq:sigma}). We obtain
$\sigma_{fs}=4.85 \sigma(M_{qh})$. From Eq.~\ref{eq:maxpdf} follows
that the maximum of $160^3$ Gaussian random numbers with variance
$\sigma_{fs}$ is distributed in the range $[23.4:27.0] \sigma(M_{qh})
$ at $90\%$ of confidence level. Combining the two $90\%$ confidence
level intervals, this means that the first PopIII progenitor of a
bright quasar originated from a perturbation in the range $[28.4:32.7]
\sigma(M_{qh})$. If we consider instead a random sub-cell among the
$160^3$, the probability that the maximum sub-grid perturbation is
smaller than $32.7 \sigma(M_{qh})$ is only $0.99995$, so several
hundreds of the $180^3$ cells among the whole simulation volume are
expected to have a PopIII progenitor formed before that of the most
massive $z=6$ halo. In fact from integration of Eq.~\ref{eq:maxpdf},
the sigma peak associated to the \emph{first} star in the box is
expected to be greater than $35.5 \sigma(M_{qh}) $ at 99.99 \% of
confidence level (and in the interval $[36.2:38.8] \sigma(M_{qh})$ at
$90\%$ of confidence level). Therefore the rarity of the earliest
PopIII progenitor of the most massive halo at $z=6$ is about $1.5
\sigma_{fs}$ less than that of the \emph{first} PopIII star formed in
the simulation volume. In terms of formation redshift, the
\emph{first} PopIII star dark matter halo in the simulation volume
virializes in the redshift interval $z \in [49:53]$, while the
earliest PopIII progenitor of the QSO halo is formed at $z \in
[38:44]$ (both intervals at 90\% of confidence level and computed for
$\sigma_8 = 0.9$).

The picture changes quite
significantly if we consider a smaller box size. E.g. in our $S1$
simulation (see Tab.~\ref{tab:sim}) with a volume of $(60 Mpc/h)^3$ a
perturbation on a mass scale $M_{qh} = 7 \cdot 10^{11} M_{\sun}/h
\approx M_{Box}/27^3$ is expected to be the most massive at $z \approx
6$. Such a halo derives at $90 \%$ of confidence level from a
fluctuation $[3.6:4.6] \sigma(M_{qh})$. If we further assume $M_{fs} =
8.5 \cdot 10^5 M_{\sun}/h = 1/94^3 M_{qh}$, we have a sub-grid
variance $\sigma_{fs}=3.66$ so that the maximum of the $94^3$ random
first light perturbation in a cell of mass $M_{qh}$ is distributed in
the range $[16.4:19.4] \sigma(M_{qh})$ at $90\%$ of confidence
level. By combining the two intervals as above, we expect that the
first PopIII progenitor of the most massive $z=6$ halo derives from a
$[20.0:24.0] \sigma(M_{qh})$ peak. The \emph{first} PopIII star
derives instead from a $[23.8:26.1] \sigma(M_{qh})$ peak (always
$90\%$ of confidence level). At variance with the larger box, here the
correlation between the most massive halo at $z=6$ and the
\emph{first} PopIII star in the simulation is expected to be stronger and the
most massive halo is likely to have as progenitor one of the first 10-100 Pop III
stars.

From these simple analytical estimates it is clear that the most
massive and rarest structures collapsed around $z \approx 6$ do not
descend from the rarest sigma peaks at the first light mass scale in
the simulation volume, when the simulation box represents a
significant fraction of the Hubble volume. Conversely the black holes
remnants of the \emph{first} PopIII stars in the universe do not
provide the seeds for super-massive black holes within the most
massive halos at $z \lesssim 6$. The descendants of first PopIII stars
are instead expected to be found at the center of a variety of halos,
as we quantify in the next Section by means of N-body simulations.

\subsection{Simulations Results}\label{sec:MCres}

In constructing the $z=6$ halo catalogs we adopt the following
parameters for the HOP halo finder \citep{eis98}. The local density
around each particle is constructed using a 16 particles smoothing
kernel. For the regrouping algorithm we use: $\delta_{peak} = 240$,
$\delta_{saddle} = 180$, $\delta_{outer} = 100$ and a minimum group
size of $20$ particles. In the large simulation box (run $L1$ in
Tab~\ref{tab:sim}) we identify $47$ halos with 20 particles or more
and the most massive halo ($37$ particles) has a mass of $6.9 \cdot
10^{12} M_{\sun}/h$. In the medium simulation box (run $M1$ in
Tab~\ref{tab:sim}) the higher mass resolution allows us to identify
$694$ halos with at least 20 particles and the most massive halo has
$92$ particles for a total mass of $6.1 \cdot 10^{12} M_{\sun}/h$,
consistent with the results from the larger box. Finally in the small
box simulations (runs $S1$ and $S2$ in Tab~\ref{tab:sim}) there are
14972 halos with at least $100$ particles in $S1$ ($\sigma_8 = 0.9$)
and 7531 halos with at least $100$ particles in $S2$ ($\sigma_8 =
0.75$). The most massive halo has a mass of $2.4\cdot 10^{12}
M_{\sun}/h$ in $S1$ and of $7.1\cdot 10^{11} M_{\sun}/h $ in $S2$. The
$z=6$ halo mass distribution for these two simulations is well
described (with displacements within $\approx 15\%$) by a
\citet{she99} mass function.

\begin{figure} 
  \plottwo{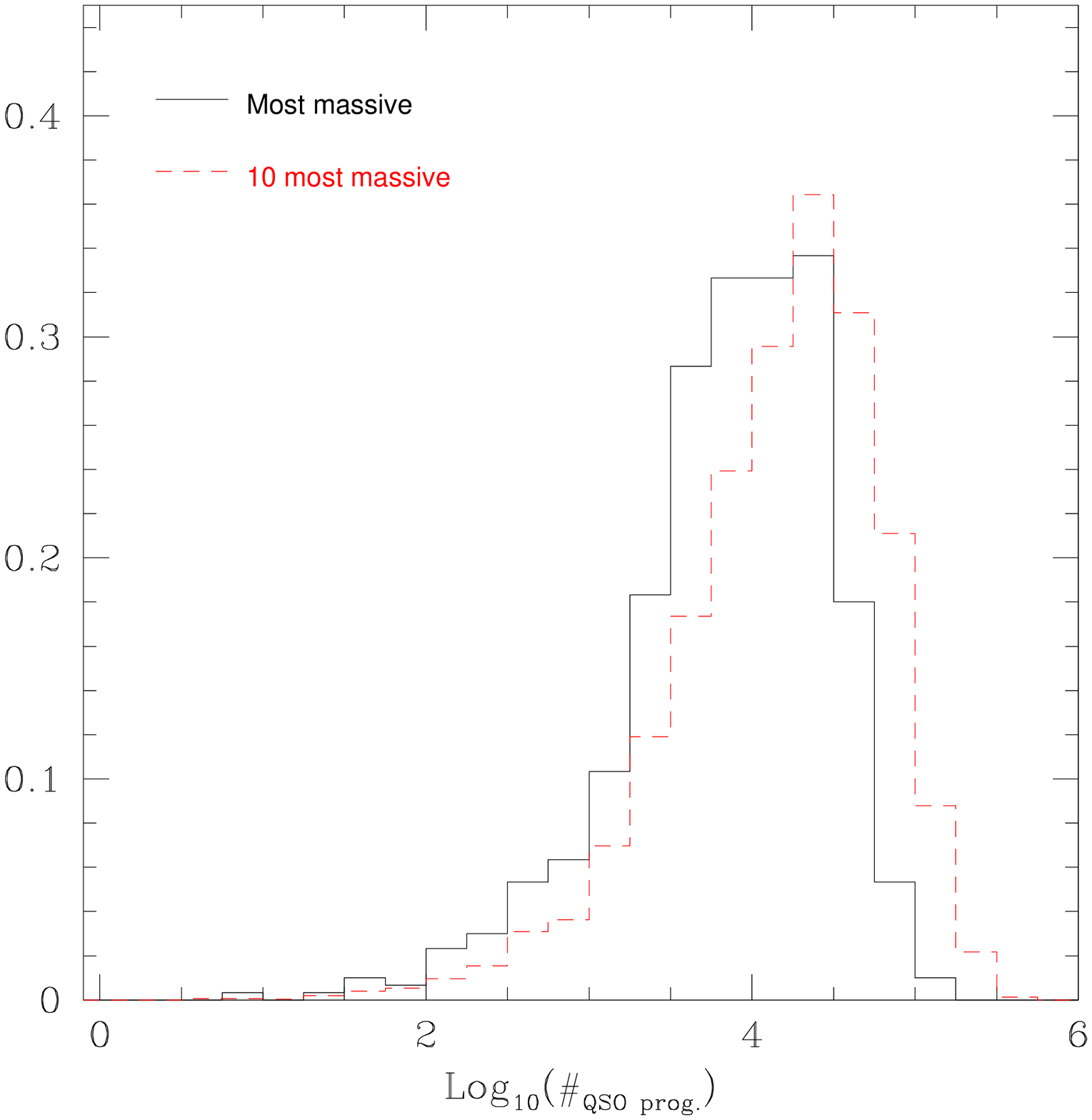}{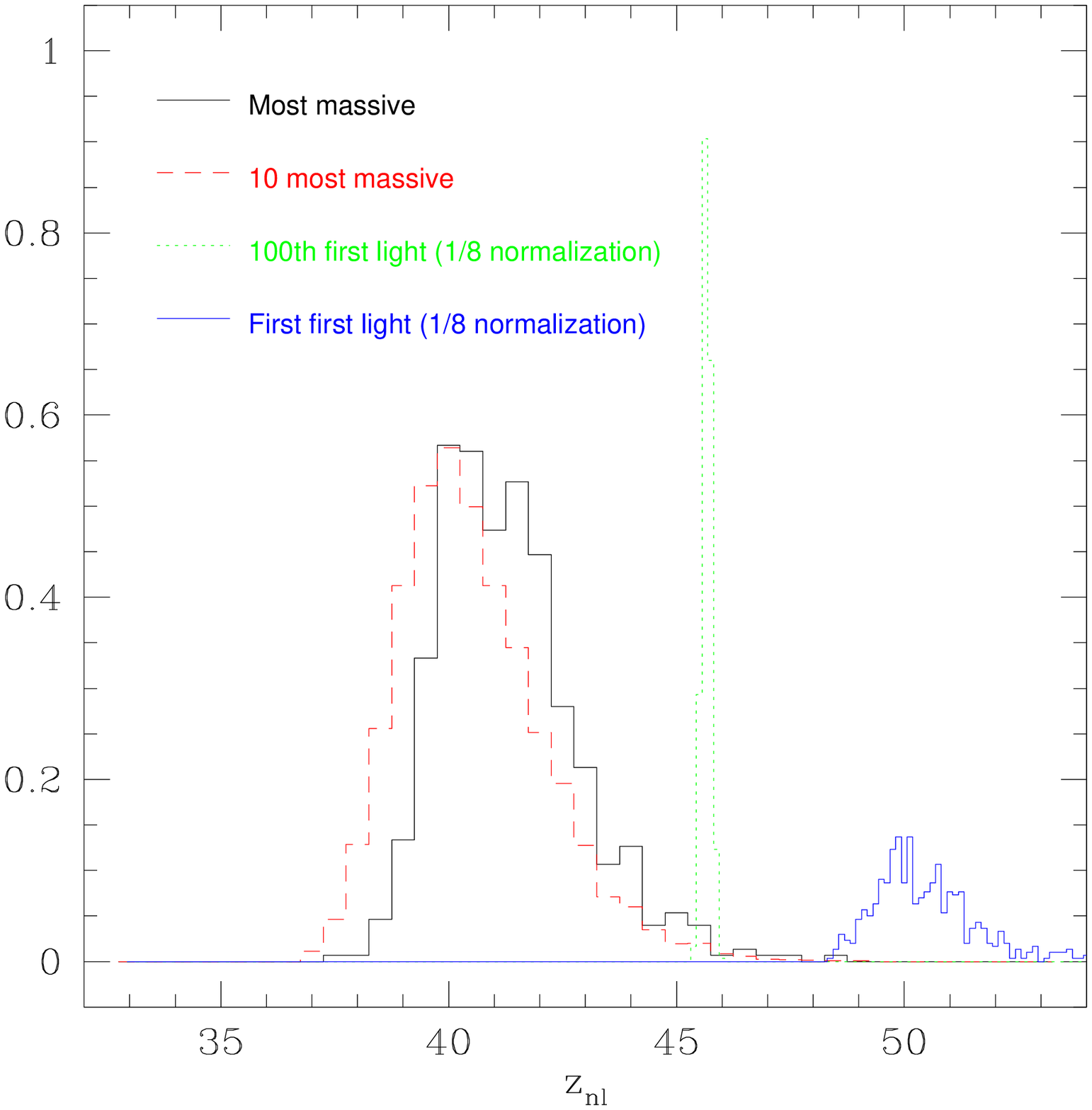} \caption{Left panel: Distribution
  for the ranking of the collapse epoch for the oldest PopIII
  halo progenitor (with $M_{fs} = 10^6 M_{\sun}/h$) of the most
  massive halo (black line) and averaged over the 10 most massive
  halos (red line) at $z=6$ in the $(720 Mpc/h)^3$ box simulation. The
  cardinality is measured over $600$ Monte Carlo realizations. Right
  panel: Like in the left panel but distribution of the collapse time
  $z_{nl}$ for the oldest PopIII progenitor. The blue line
  represents the collapse redshift of the \emph{first} PopIII star
  perturbation, while the dotted green line refers to the collapse
  redshift of the 100th PopIII in the box.}\label{fig:card720}
\end{figure}

\begin{figure} 
  \plotone{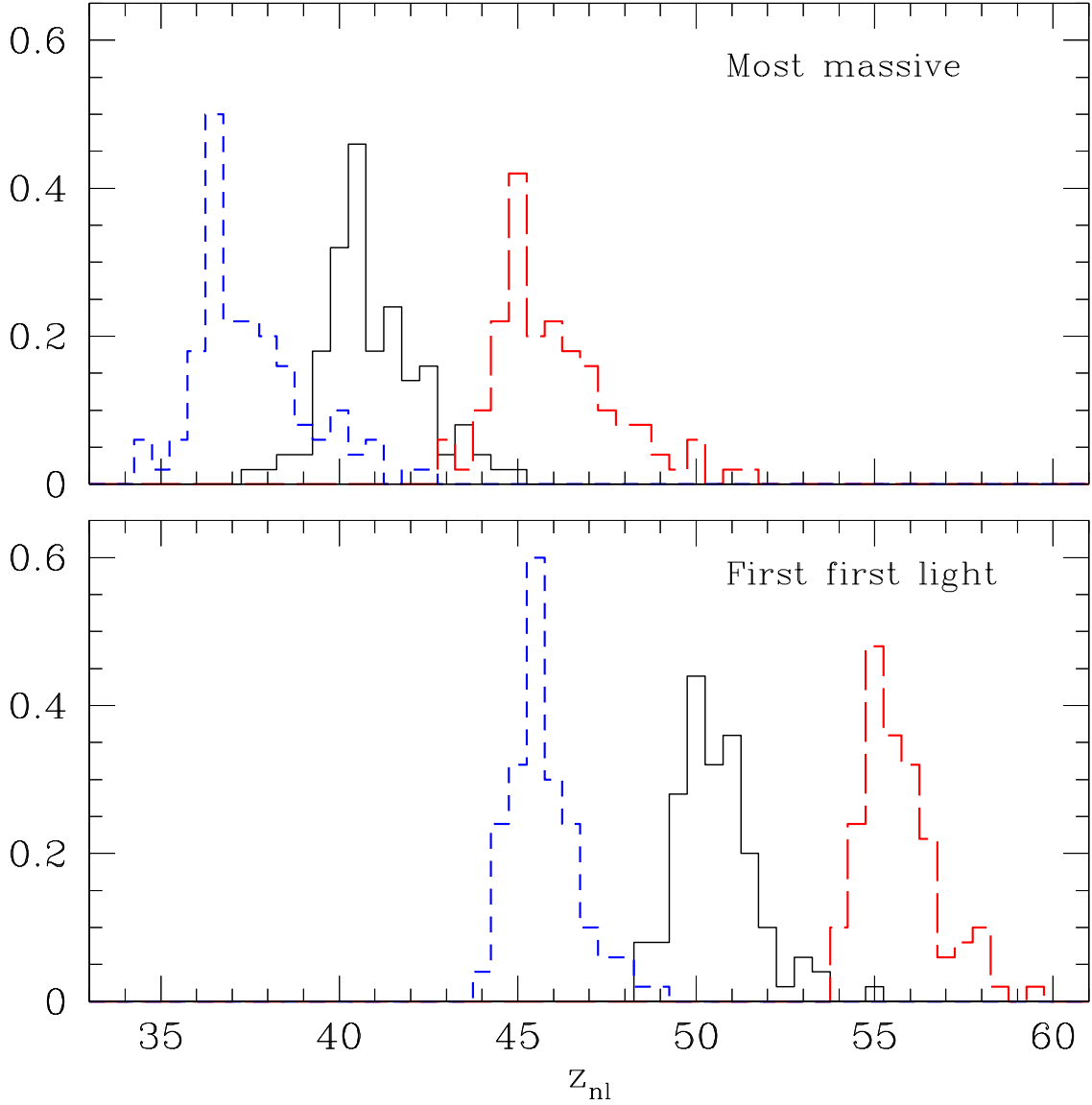}\caption{Probability distribution of the collapse
  time for the oldest PopIII progenitor of the most massive halo at
  $z=6$ (upper panel) and for the first PopIII halo formed in the box
  (bottom panel) in the $(720 Mpc/h)^3$ box simulation. The curves
  refer to different masses for the PopIII halo: $M_{fs}= 3.4 \cdot
  10^6 M_{\sun}/h$ (blue, short dashed), $M_{fs}= 1.0 \cdot 10^6
  M_{\sun}/h$ (black, solid) and $M_{fs}= 3.0 \cdot 10^5 M_{\sun}/h$
  (red, long dashed).  }\label{fig:mass}
\end{figure}

The link between the halos identified in the snapshots and the first
light sources is established using the Monte Carlo method described in
Sec.~\ref{sec:MC}. For the large box we consider a refinement factor
$k=57^3$ to move from the single particle mass of the simulation to a
typical PopIII halo mass, so that $M_{fs} = 1.0 \cdot 10^{6}
M_{\sun}/h$. For the first 10 most massive halos at $z=6$ we show in
Fig.~\ref{fig:card720} the distribution of the redshift at which the
oldest progenitor crosses the virialization density contrast threshold
in linear theory ($\delta \rho /\rho = 1.69$) and the distribution of
the ranking of the collapse time computed over all the PopIII
progenitors of the simulation particles. The collapse rank of the
first PopIII progenitor of the most massive $z=6$ halo is in the
interval $[474:45075]$ at $90\%$ of confidence level, with median
$8535$. The corresponding virialization redshifts are in the interval
$[39.3:44.1]$ with median $41.1$. For comparison the \emph{first}
PopIII halo in the box virializes in the redshift interval
$[49.0:52.7]$ with median $50.3$; the 100th first light in the box
collapses in the redshift range $[45.5:45.8]$ with median
$45.7$. These results from the combined N-body simulation and Monte
Carlo code are in excellent quantitative agreement with the analytical
estimates of Sec.~\ref{sec:est} and confirm that in a large simulation
box the most massive halos at $z=6$ do not derive from the rarest
sigma peaks at the first light mass scale. This result is robust with
respect to the adopted typical mass for PopIII halos. In
Fig.~\ref{fig:mass} we show the results obtained by changing the mass
of the halos hosting the first stars considering larger halos ($M_{fs}
= 3.4 \cdot 10^6 M_{\sun/h}$ with $k=38^3$) and smaller halos ($M_{fs}
= 3.0 \cdot 10^5 M_{\sun/h}$ with $k=85^3$). The formation redshift
varies as the first halos are formed earlier when they are less
massive, but the relative ranking between the first PopIII halo in the
box and the first PopIII progenitor of the most massive structures at
$z=6$ remains similar. In passing we note that our distribution of the
formation redshift for the first $3 \cdot 10^{5} M_{\sun}$ progenitor
of the most massive $z=6$ halo (formed at $z \approx 46$) is in
agreement with the results by \citet{ree05}, obtained by means of
N-body simulations with adaptive refinements. However this halo is not
the first one formed in the simulation box as we find that the first
structure on this mass scale is formed at $z \gtrsim 55$ (see
Fig.~\ref{fig:mass}).

\begin{figure} 
  \plottwo{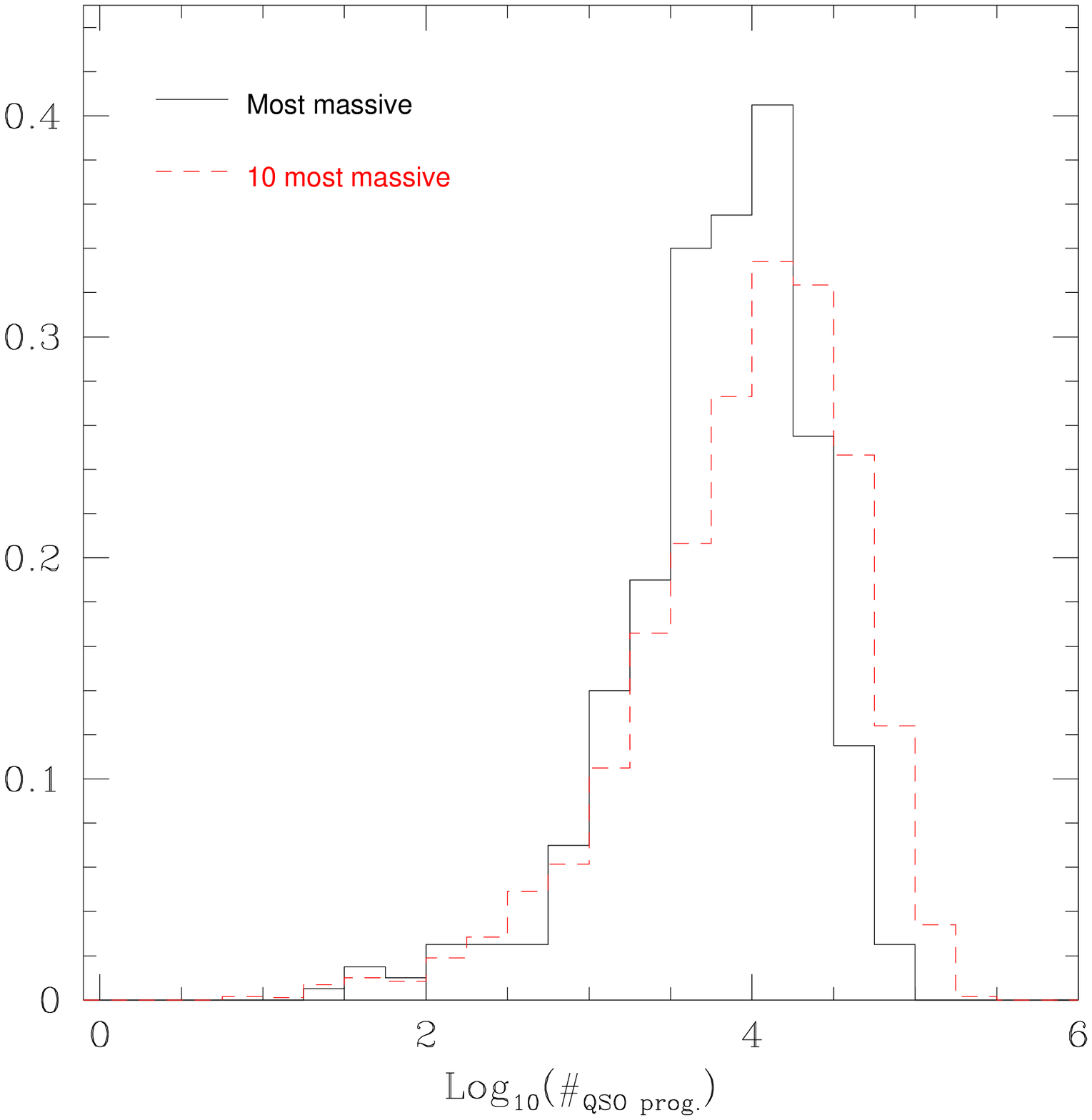}{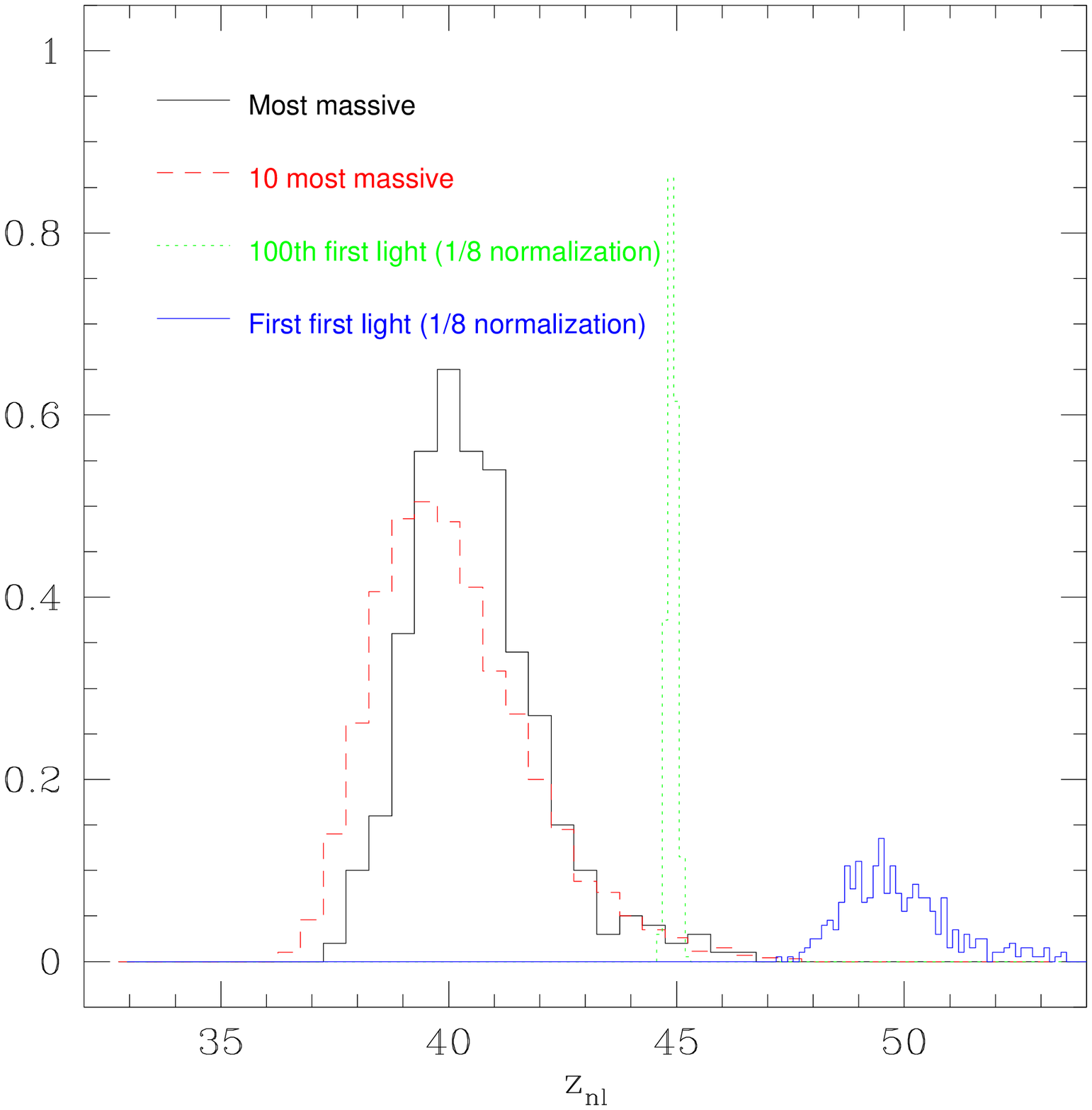} \caption{Like in
  Fig.~\ref{fig:card720}, but for a simulation of a $(512
  Mpc/h)^3$ volume and using 200 Monte Carlo realizations.
  }\label{fig:card512}
\end{figure}

The results are similar for the medium box, which has a volume that is
only three times smaller than the large one (see
Fig.~\ref{fig:card512}). The refinement factor used here is $k=40$
that gives $M_{fs} = 1.0 \cdot 10^{6} M_{\sun}/h$. The \emph{first}
PopIII halo in the box virializes in the redshift range $[48.3:52.4]$
with median $49.7$, while the oldest PopIII progenitor of the most
massive halo virializes in the redshift range $[38.5:43.2]$ (median
$40.4$) and has a collapse ranking in $[574:37499]$ at $90\%$ of
confidence level, with median $7261$.

\begin{figure} 
  \plottwo{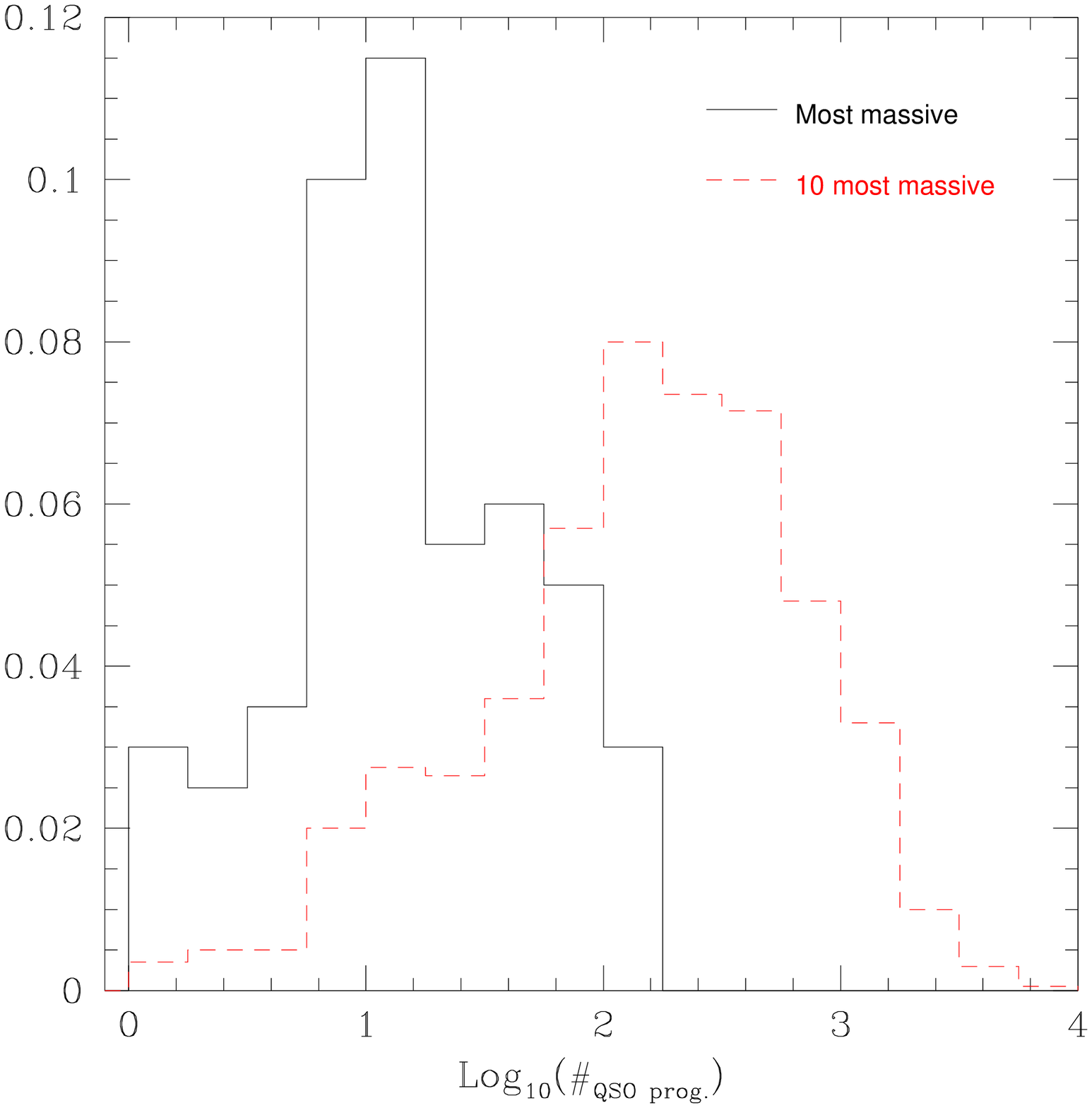}{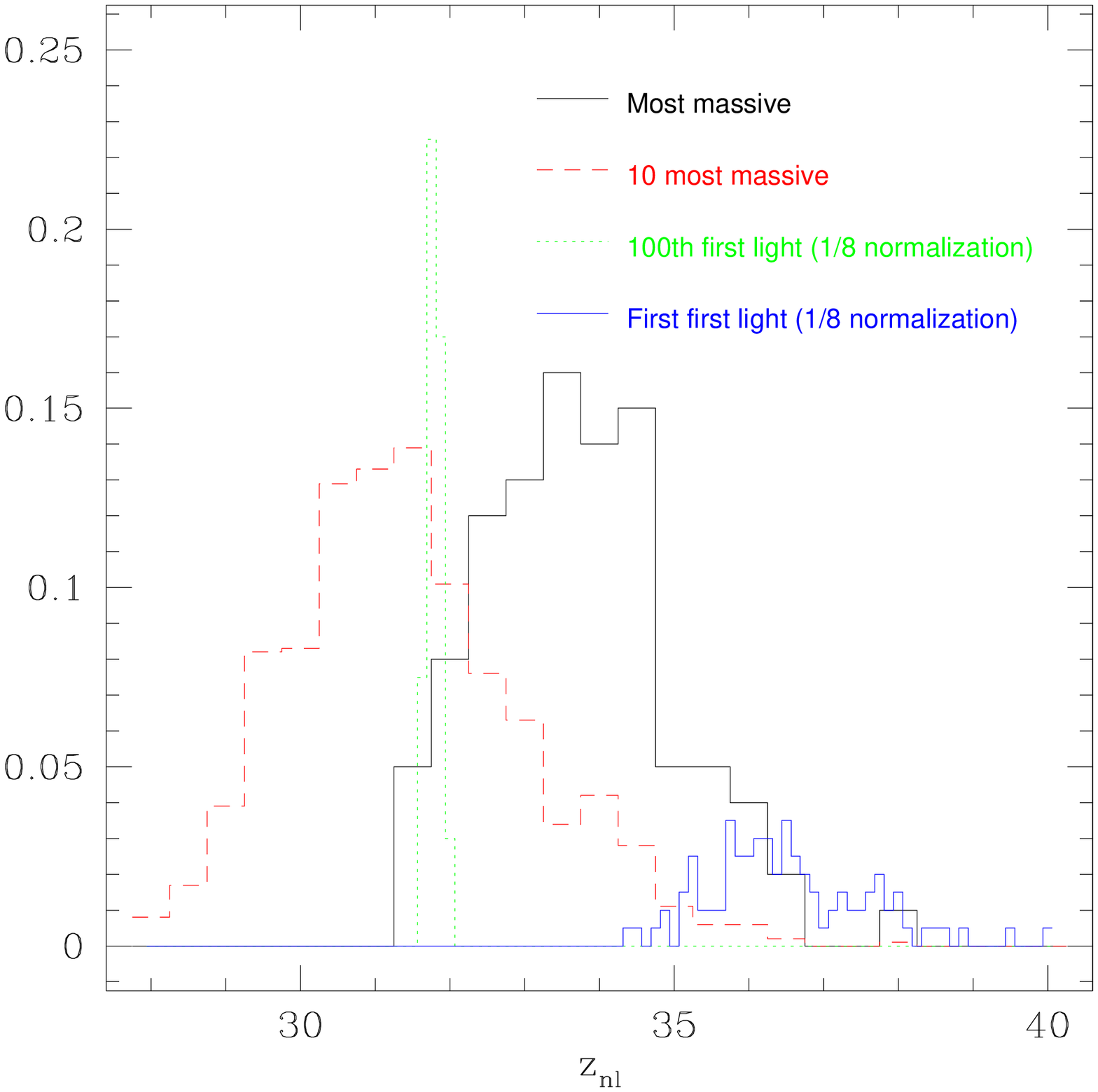} \caption{Like in
  Fig.~\ref{fig:card720}, but for the $S2$ simulation, with volume of
  $(60 Mpc/h)^3$ and $\sigma_8 = 0.75$}\label{fig:card60}
\end{figure}

The picture changes significantly (see Fig.~\ref{fig:card60}) for the
small box that has a volume more than $10^3$ times smaller than the
large box. Here we use a refinement factor $k=5$, that leads to
$M_{fs} = 8.6 \cdot 10^{5} M_{\sun}/h$. The collapse rank of the first
light progenitor of the most massive $z=6$ halo is in the range
$[1:103]$ at the $90 \%$ confidence level with median $13$. The
correlation between the \emph{first} PopIII star and the most massive
structure at $z=6$ is therefore strong due to the small volume of the
box. This means that \emph{locally} the oldest remnants of first stars
are expected to be within the largest collapsed structures.

From the medium box size numerical simulation we have also
characterized the fraction of \emph{first} PopIII remnants that end up
in identified halos at $z=0$. If we consider one of the first 100
first light halos collapsed in the box, there is an average
probability of $0.72$ of finding its remnant in a halo identified at
$z = 0$ with more than 100 particles (that is of mass above $6.7 \cdot
10^{12} M_{\sun}/h$). The median distribution for the mass of a halo
hosting one of the remnants of these first light sources is $\approx 3
\cdot 10^{13} M_{\sun}/h$. At $95\%$ of confidence level the remnants
are hosted by a halo of mass less than $3.6\cdot 10^{14}
M_{\sun}/h$. For comparison, the most massive halo in the simulation
has a mass of $ 4.4 \cdot 10^{16} M_{\sun}/h$ and there are about
$15000$ halos more massive than $ 3 \cdot 10^{12} M_{\sun}/h$.  This
is a consequence of the poor correlation between first PopIII halos
and most massive halos at low redshift.

\begin{figure} 
  \plotone{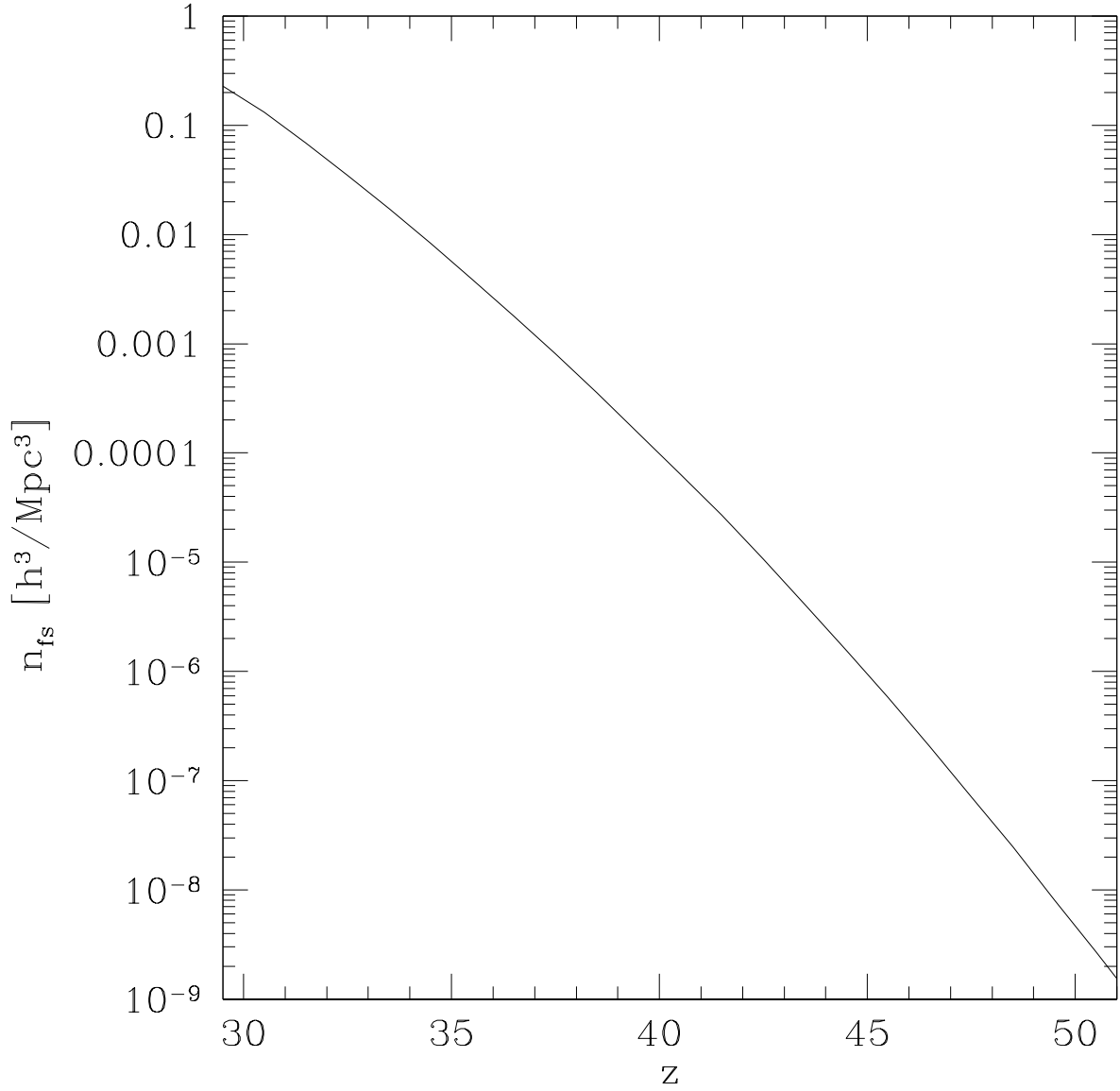}\caption{Cumulative comoving number density of
  virialized PopIII halos ($M_{fs}=10^6 M_{\sun}/h$) versus redshift for
  $z>29$ as measured by means of our MC code.}\label{fig:sfr}
\end{figure}

Finally, combining the results from all our three simulation boxes, we
construct in fig.~\ref{fig:sfr} the PopIII star formation rate at high
$z$. The total number of PopIII halos that virialize is increasing
with redshift, reaching a number density of $\approx 0.1 (Mpc/h)^{-3}$
at $z\approx 30$. In our small box simulation, this means that the
average density of PopIII halos is $\approx 10^{-3}$ per grid
cell. Therefore there is a very small probability of having two
collapsed halos within the same cell, an event that would not be
captured in our model. 

\section{When does the first stars epoch end?} \label{sec:end_of_fs}

In Sec.~\ref{sec:first_halos} we show that the most massive halos at
$z=6$ have first light progenitors that have been formed when already
several thousands of other PopIII stars existed. Are these progenitors
still entitled to be called \emph{first stars}?  That is, when does
the \emph{first stars} epoch end? Here we review the question adopting
two different definitions to characterize the transition from the
first to the second generation of stars, namely (i) a threshold for
the transition given by the destruction of molecular hydrogen and (ii)
a metallicity based threshold.

\subsection{Molecular Hydrogen destruction} \label{sec:h2}

One criterion for the end of the first light epoch can be based on the
destruction of Molecular Hydrogen in the ISM due to photons in the
Lyman-Werner ($[11.15:13.6] eV$) energy range emitted by PopIII
stars. $H_2$ is in fact needed for cooling of the gas in dark matter
halos of mass $\approx 10^6$ \citep[e.g. see][]{bro04}. The flux in
the Lyman-Werner band is about $7.5\%$ of the ionizing flux (i.e. with
an energy range above $13.6 eV$). A PopIII star is expected to emit a
total of about $7.6 \cdot 10^{61}$ photons per solar mass
\citep{sti04}, so if we assume $300 M_{\sun}$ as a typical mass we
have about $1.7 \cdot 10^{63}$ $H_2$-destroying photons emitted over
the stellar lifetime. Only a fraction $\approx 0.15$ of these photons
can effectively destroy an $H_2$, molecule, as the most probable
outcome of absorption of a Lyman Werner photon is a first decay to a
highly excited a vibrational level that later returns to the
fundamental state, with resulting re-emitted photons below the $11.15$
eV threshold \citep{shu82,glo01}. Therefore we estimate that $\approx
2.5 \cdot 10^{62}$ $H_2$ molecules will be destroyed by a PopIII
star. Given the neutral hydrogen number density $6.2 \cdot 10^{66}
Mpc^{-3}$ this means that a PopIII star destroys $H_2$ over a volume
$4 \cdot 10^{-5} Mpc^{3} /\xi $, where $\xi$ is the ratio of molecular
to atomic hydrogen. Assuming a primordial molecular hydrogen fraction
$\xi \approx 10^{-6}$ (e.g. see \citealt{peebles}), we obtain that a
PopIII star has the energy to destroy primordial $H_2$ in a volume of
$\approx 14 (Mpc/h)^3$. This number is in broad agreement with
detailed radiative transfer simulations by \citet{joh07b}. From
Fig.~\ref{fig:sfr}, it is immediate to see that by $z\approx 30$ the
PopIII number density has reached the critical level of $0.1
(Mpc/h)^{-3}$ and therefore around that epoch the radiation background
destroys all the primordial $H_2$.  Once all the primordial $H_2$ has
been cleared the universe becomes transparent in the Lyman Werner
bands and the new $H_2$ formed during the collapse of gas clouds is
dissociated by the background radiation. In fact, assuming that the
abundance of $H_2$ formed during collapse is $\xi_{coll}\approx 5
\cdot 10^{-4}$ (e.g. see \citealt{hai00}), this means that a
collapsing $10^6M_{\sun}$ halo produces about $1.3\cdot 10^{59}$ $H_2$
molecules, a negligible number with respect to the $2.5\cdot 10^{62}$
$H_2$ that are destroyed. Our simple estimate therefore suggests that
around $z\approx 30$ the star formation rate of PopIII stars in
$10^{6} M_{\sun}$ halos is greatly suppressed and proceeds in a
self-regulated fashion where only a fraction $\approx 10^{-3}$ of the
collapsing halos are actually able to cool and lead to the formation
of massive PopIII stars. Eventually the Lyman Werner background is
maintained by PopIII stars formed in more massive halos ($M \approx
10^8 M_{\sun}$), cooled by atomic hydrogen, and, at later times, by
PopII stars.

Inspired by these ideas we set the end of the \emph{primordial} epoch
for PopIII formation at the point where the primordial $H_2$ has been
destroyed, that is around $z\approx 30$. Of course this is only an
order of magnitude estimate and to fully address the feedback due to
photo-dissociating Lyman Werner photons realistic radiative transfer
cosmological simulations are needed, which may led even to positive
feedback \citep[e.g. see][]{ric01}. In particular our estimate does
not take into account the effects of self-shielding and the fact that
the formation timescale for $H_2$ during the halo collapse may be
faster than the timescale for photo-dissociation by the background
radiation. Thus it is possible that the PopIII star formation rate at
$z \lesssim 30$ is not suppressed as much as predicted by our
argument. However our estimate seems to be in broad agreement with the
more realistic model by \citet{hai00} that predicts the onset of a
significant negative feedback at $z \gtrsim 25$, depending on the
assumed efficiency of Lyman Werner photon production.

\subsection{Metal enrichment}

Another possibility to end the first star epoch is based on a ISM
metallicity threshold. However, in this case a clear transition epoch
is missing (e.g. see \citealt{sca03,fur05}). This is because metal
enrichment, driven by stellar winds whose typical velocities are many
orders of magnitude lower than the speed of light, is mainly a local
process. Therefore pockets of primordial gas may exists in regions of
space that have experienced relatively little star formation, such as
voids, even when the average metallicity in the universe is above the
critical threshold assumed to define the end of the PopIII era.

In any case this definition provides a longer duration for the first
star era. In fact to enrich the local metallicity above the $Z =
10^{-4} Z_{\sun}$ threshold, relevant for stopping PopIII formation by
chemical feedback (see \citealt{bro04}), one $300 M_{\sun}$ SN must
explode for every $\approx 2\cdot 10^{8} M_{\sun}$ total mass volume
(DM+barions), assuming on average a PopIII mass of $300 M_{\sun}$ with yield
$0.2$. For a Milky Way like halo, this means that about 3000 first
stars SN are needed to enrich the IGM to the critical
metallicity. Accordingly to this definition, the Pop III epoch would
end within a significant fraction of the total simulated volume around
$z \approx 20$, when there is the collapse of dark matter halos
originated from $3\sigma$ peaks at the $10^{6} M_{\sun}$ mass scale
\citep[e.g., see][]{mad01}, if the suppression in the PopIII star
formation rate due to lack of $H_2$ cooling is neglected. A further
caveat is that very massive stars may end up directly in Intermediate
Mass Black Holes without releasing the produced metals in the IGM
\citep{meg02,san02}.

\section{Growth of the PopIII black hole seeds} \label{sec:BHgrow}

From our investigation it is clear that, before the first PopIII
progenitor of the most massive halo at $z=6$ is born, several
thousands of intermediate mass ($m_{BH}\approx 10^2{M_{\sun}}$) black
hole seeds are planted by PopIII stars formed in a cosmic volume that
will on average host a bright $z=6$ quasar. This result does not allow
to establish an immediate correlation between the very first PopIII
stars created in the universe and the bright $z=6$ quasars, but
neither does it exclude such a link, as the formation epoch for the
quasar seed is still at very high redshift ($z \gtrsim 40$), when
radiative feedback from other PopIII stars already formed is unlikely
to affect the formation and evolution of the seed (see
Sec.~\ref{sec:h2}). Here we investigate with a simple merger tree code
what is the fate of the black holes seeds formed up to the formation
time of the quasar seed and what are the implications for the observed
quasar luminosity function.

We assume Eddington accretion for the BH seeds, so that the evolution
of the BH mass is given by: 
\begin{equation}
m_{BH} = m_{0} \exp{\left [(t-t_0)/t_{sal} \right ]},
\end{equation}
where $m_0$ is the mass at formation time $t_0$ and $t_{sal}$ is the
Salpeter time \citep{sal64}:
\begin{equation} \label{eq:accr}
t_{sal} = \frac{\epsilon ~m_{BH} ~c^2}{(1-\epsilon)L_{Edd}} = 4.507 \cdot
10^{8} yr \frac{\epsilon}{(1-\epsilon)}, 
\end{equation}
where $\epsilon$ is the radiative efficiency.

Using Eq.~\ref{eq:accr} we can immediately see that a difference of
$\Delta t \lesssim 2 \cdot 10^7 yr$, that is of $\Delta z \approx 10$
at $z=40$, in the formation epoch of the BH seed of a bright quasar is
not too important in terms of the final mass that can be accreted by
$z=6$, as this corresponds to about half a folding time. Assuming $\epsilon
= 0.1$ until $z=6.4$, the highest redshift in the SDSS quasar sample
\citep{fan04}, we obtain a ratio of final to initial mass
$m_{BH}/m_0 = 2.62 \cdot 10^7$ for $z=50$ and $m_{BH}/m_0 = 1.78
\cdot 10^7$ for $z=40$. Therefore in both cases there has been enough
time to build up a $z\approx 6$ supermassive black hole with mass
$m_{BH} \gtrsim 10^9$ starting from a PopIII remnant.

This estimate however highlights that only a minor fraction of the
PopIII BH seeds formed before $z=40$ can accrue mass with high
efficiency, otherwise the number density of supermassive black holes
at low redshift would greatly exceed the observational
constraints. The first BH seeds in the box are distant from each
other, so they evolve in relative isolation, without possibly merging
among themselves. Therefore other mechanisms must be responsible for
quenching accretion of the first BH seeds. Interestingly if we were to
assume that accretion periods are Poisson distributed in time for each
seed, we would not be able to explain the observed power law
distribution of BH masses at $z\lesssim 6$ around the high mass end. A
Poisson distribution would in fact give too little scatter around the
median value and a sharp (faster than exponential) decay of the
displacements from the mean accreted mass. An exponential distribution
of the accretion efficiency is instead required to match the observed
BH mass function. In addition, it is necessary to assume that the duty
cycle of the BH accretion is roughly proportional to the mass of the
halo it resides in. This is sensible, since an accretion model
unrelated to the hosting halo mass may lead to the unphysical result
of possibly accruing more mass than the total baryon mass available
in that halo. In fact, a BH seed formed at $z\approx 40$ is within a
halo of median mass $\approx 10^{11} M_{\sun}/h$ at $z=6$ and a few
percent of the seeds may be in halos with mass below $\approx 10^{9}
M_{\sun}/h$ at that redshift.

\begin{figure} 
  \plotone{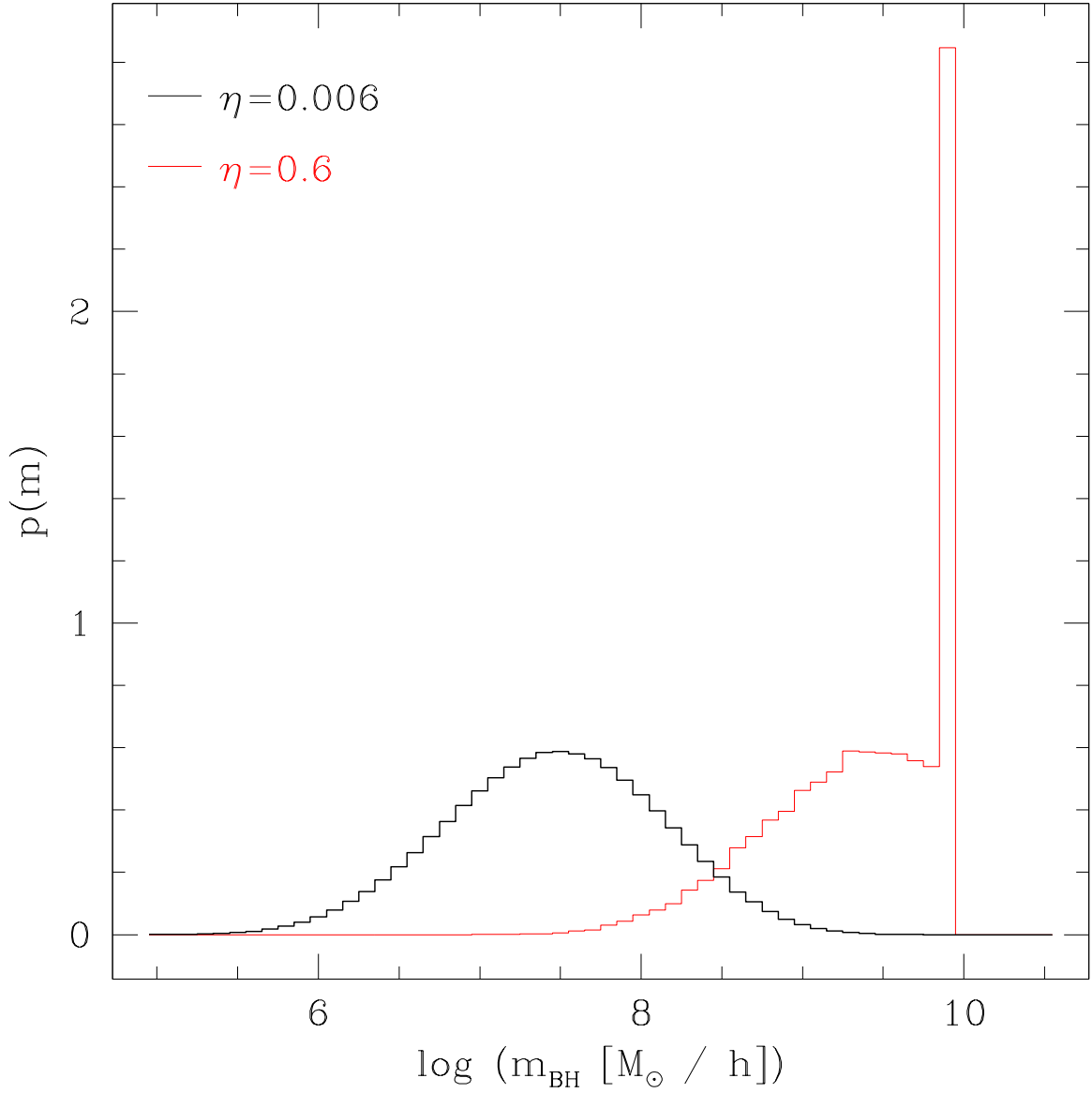}\caption{Final BH mass at $z=6$ for a $100
  M_{\sun}/h$ seed starting Eddington accretion with radiative
  efficiency $\epsilon=0.1$ at $z=40$. The growth of the BH is limited
  to a fraction $\eta$ of the total baryon mass of its halo, as
  obtained with a merger-tree Monte Carlo code.}\label{fig:accr}
\end{figure}

To explore this possibility we follow the merging history of PopIII
halos formed at $z=40$ by means of a merger-tree code based on
\citet{lac93}. We implement a BH growth based on Eq.~\ref{eq:accr},
but at each step of the tree we limit the BH mass to $m_{BH} \leq
\eta~ m_{bar}$, where $m_{bar}$ is the total baryon mass of the halo
that hosts the BH. The results are reported in fig.~\ref{fig:accr}. If
the BH growth is not constrained (or only mildly constrained), then a
significant fraction of the seeds grows above $10^{10} M_{\sun}$,
which would result in an unrealistic number density of supermassive
black holes at $z=6$. However, if $\eta \approx 6 \cdot 10^{-3}$ (like
in \citealt{yoo04}; see also \citealt{wyi03}), then we obtain an
expected mass for the BH powering bright $z=6$ quasar of $\approx 5
\cdot 10^{9} M_{\sun}$, which is in agreement with the observational
constraints from SDSS quasars \citep{fan04}. By fitting a power law
function to the BH mass function in the range $[0.055:0.2]\cdot
10^{10} M_{\sun}$ we obtain a slope $\alpha \approx -2.6$, while the
slope is $\alpha \approx -3.7$ in the mass range $[0.2:1.0] \cdot
10^{10} M_{\sun}$, a value that is consistent within the $1 \sigma$
error bar with the slope of the bright end of the quasar luminosity
function measured by \citet{fan04}.

Another contribution to ease an overproduction of bright quasars may
be given by the suppression of the early growth of the PopIII BH seeds
for the first $\approx 10^8 yr$ after formation, that is for about $2
t_{sal}$ \citep{joh07}. In fact the radiation from a PopIII star may
evacuate most of the gas from its host halo, so that the subsequent BH
growth is quenched until a merger provides a new gas reservoir to
enable growth at near Eddington rate \citep{joh07}. Also the BH seeds
situated in more massive halos would probably be more likely to
replentish their gas supply earlier.

\section{Conclusion} \label{sec:conc}

In this paper we investigate the link between the first PopIII halos
collapsed in a simulation box and the most massive structures at $z
\approx 6$, with the aim of establishing the relationship between the
first intermediate mass black holes created in the universe and the
super-massive black holes that power the emission of bright $z=6$
quasars. We show that almost no correlation is present between the
sites of formation of the first few hundred $10^6 M_{\sun}/h$ halos
and the most massive halos at $z \lesssim 6$ when the simulation box
has an edge of several hundred $Mpc$. Here the PopIII progenitors
(halos of mass $M_{fs} \approx 10^6 M_{\sun}$) of massive halos at $z
\lesssim 6$ formed from density peaks that are $\approx 1.5
\sigma(M_{fs})$ more common than that of the \emph{first} PopIII star
in the $(512 Mpc/h)^3$ simulation box. These halos virialize around
$z_{nl} \approx 40$, to be compared with $z_{nl} \gtrsim 48$ of the
\emph{first} PopIII halo.

This result has important consequences. We show that, if bright
quasars and supermassive black holes live in the most massive halos at
$z\approx 6$, then their progenitors at the $10^6 M_{\sun}$ mass scale
are well within the PopIII era, regardless of the PopIII termination
mechanism. On the other hand, if the $m_{BH}/\sigma$ relationship is
already in place at $z=6$, then bright quasars are not linked to the
remnants of the very first intermediate mass black holes (IMBHs) born
in the universe, as their IMBH progenitors form when already several
thousands of PopIII stars have been created within the typical volume
that hosts a bright $z=6$ quasar. The IMBH seeds planted by this very
first PopIII stars have sufficient time to grow up to $m_{BH} \in
[0.2:1] \cdot 10^{10} M_{\sun}$ by $z=6$ if we assume Eddington
accretion with radiative efficiency $\epsilon \lesssim 0.1$. Instead,
quenching of the BH accretion is required for the seeds of those
PopIII stars that will not end up in massive halos at $z=6$, otherwise
the number density of supermassive black holes would greatly exceed
the observational constraints. One way to obtain growth consistent
with the observations is to limit the accreted mass at a fraction
$\eta \approx 6 \cdot 10^{-3}$ of the total baryon halo mass. This
gives a slope of the BH mass function $\alpha = -3.7$ in the BH mass
range $m_{BH} \in [0.2:1] \cdot 10^{10} M_{\sun}$, which is within the
$1 \sigma$ uncertainty of the slope of the bright end of the $z=6$
quasar luminosity function ($\alpha \approx -3.5$) measured by
\citet{fan04}.

Another important point highlighted by this study is that rich
clusters do not preferentially host the remnants of the first PopIII
stars. In fact the remnants of the first 100 Pop-III stars in our
medium sized simulation box (volume of $(512 Mpc/h)^3$) end up at
$z=0$ on halos that have a median mass of $3 \cdot 10^{13}
M_{\sun}/h$. This suggests caution in interpreting the results from
studies that select a specific volume of the simulation box, like a
rich cluster, and then progressively refine smaller and smaller
regions with the aim of hunting for the first lights formed in the
whole simulation \citep[see e.g., ][]{ree05}. Only by considering
refinements over the complete volume of the box the rarity and the
formation ranking of these progenitors can be correctly evaluated.

\acknowledgements

We thank Mike Santos for sharing his code to generate the initial
conditions and for a number of useful and interesting discussions. We
are grateful to the referee for constructive suggestions. This work
was supported in part by NASA JWST IDS grant NAG5-12458 and by
STScI-DDRF award D0001.82365. This material is based in part upon work
supported by the National Science Foundation under the following NSF
programs: Partnerships for Advanced Computational Infrastructure,
Distributed Terascale facility (DTF) and Terascale Extensions:
enhancements to the Extensible Terascale Facility - Grant AST060032T.
 

\appendix
\section{Tests for the First Light Monte Carlo Method}\label{sec:app}

\begin{figure} 
  \plotone{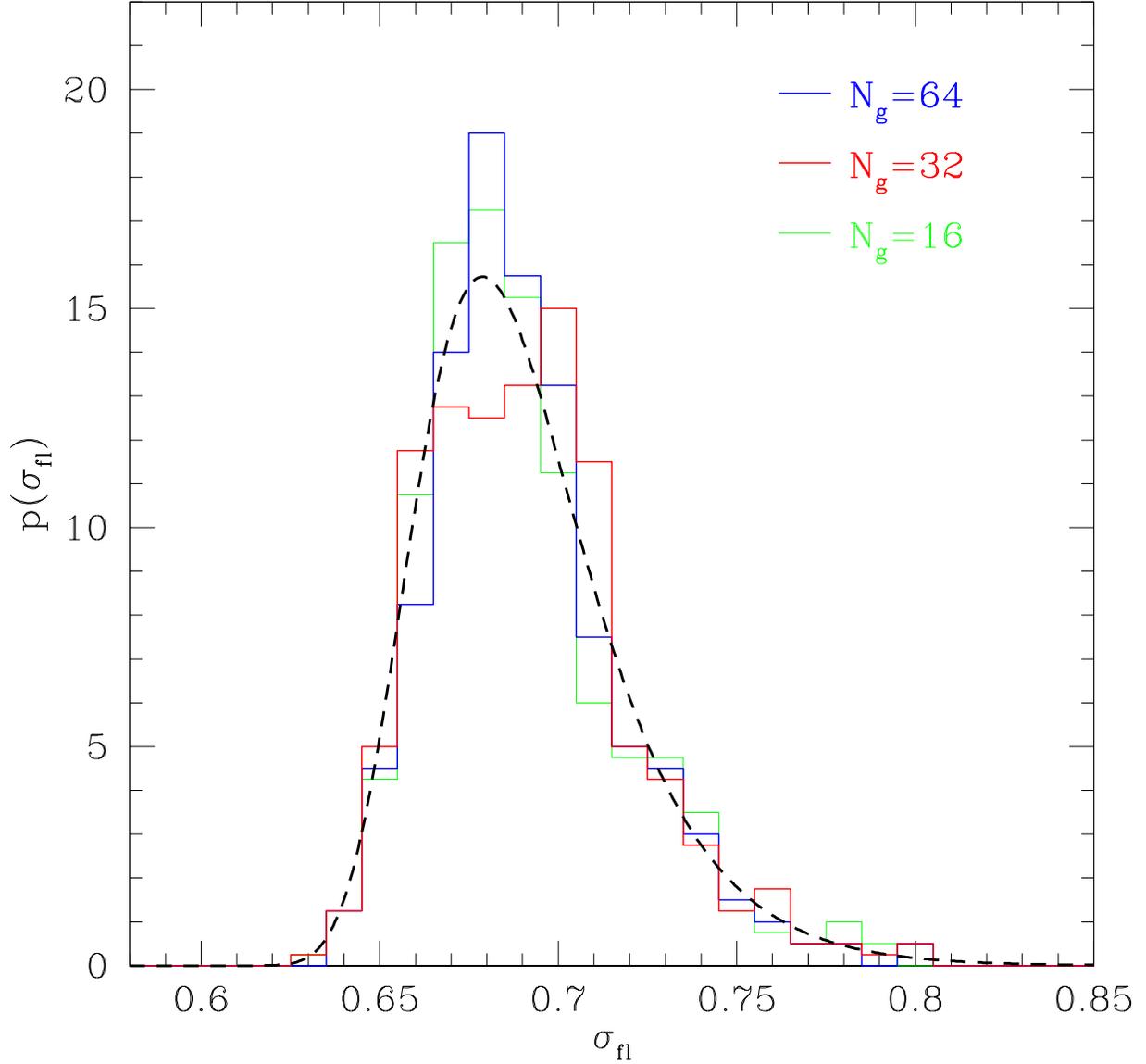} \caption{Probability distribution for the maximum
  density fluctuation at first light scale ($\sigma_{fl}$) in a
  synthetic test of the refinement method obtained generating first a
  gaussian random field with $\sigma=0.1$ on a $N_g^3=64^3$ grid and
  then applying our MC method with a refinement factor $4$, subgrid
  fluctuations $\sigma_{fs}=0.08$ and 400 Monte Carlo realizations. We
  progressively downgrade the resolution of the simulation grid to
  $N_g=32$ and $N_g=16$, increasing $\sigma_{fs}$ following our
  prescription. The MC sampling results are compared with the
  theoretical probability distribution for $\sigma_{fl}$ (dashed
  line). }\label{fig:check}
\end{figure}

\begin{figure} 
  \plottwo{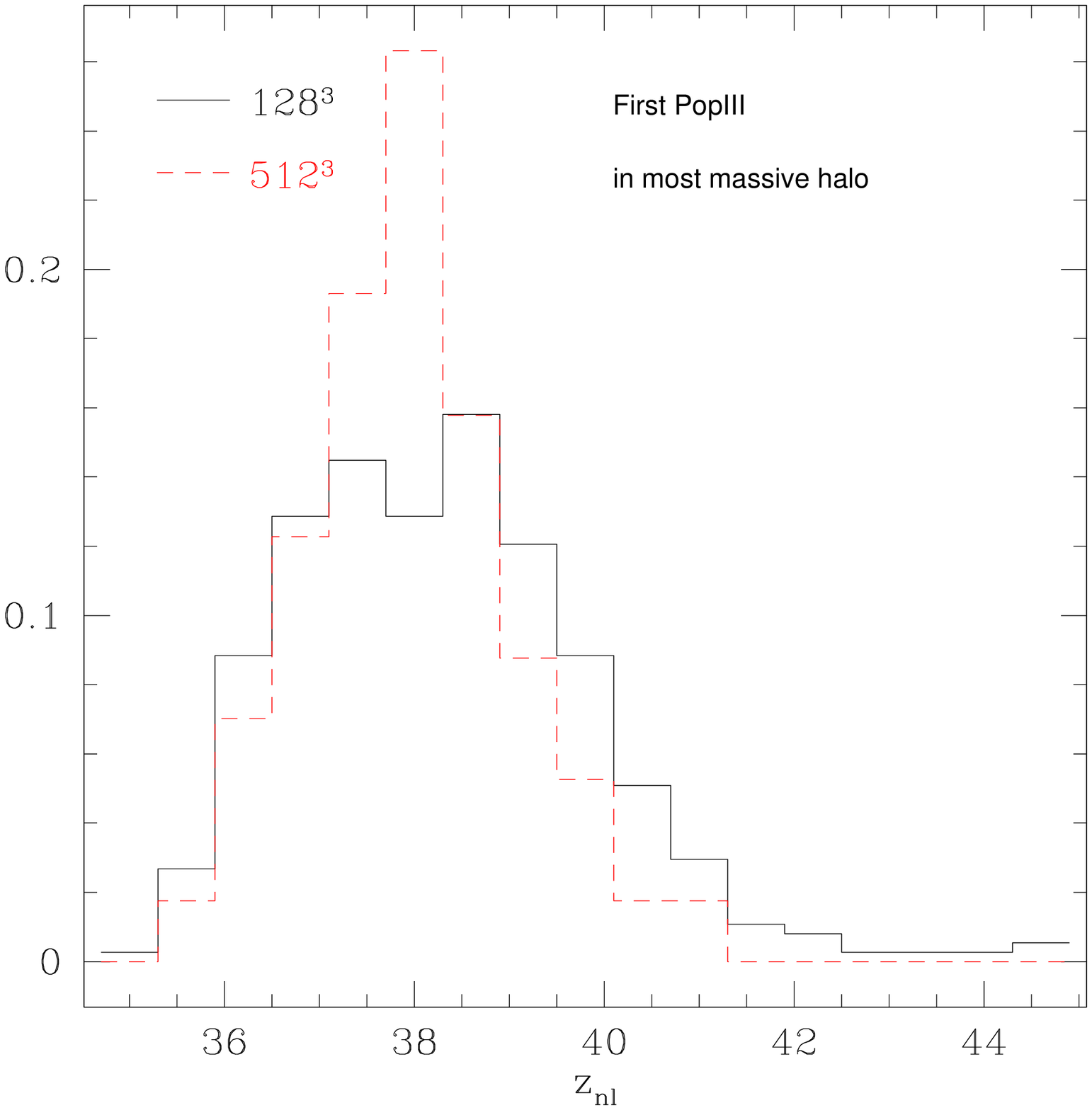}{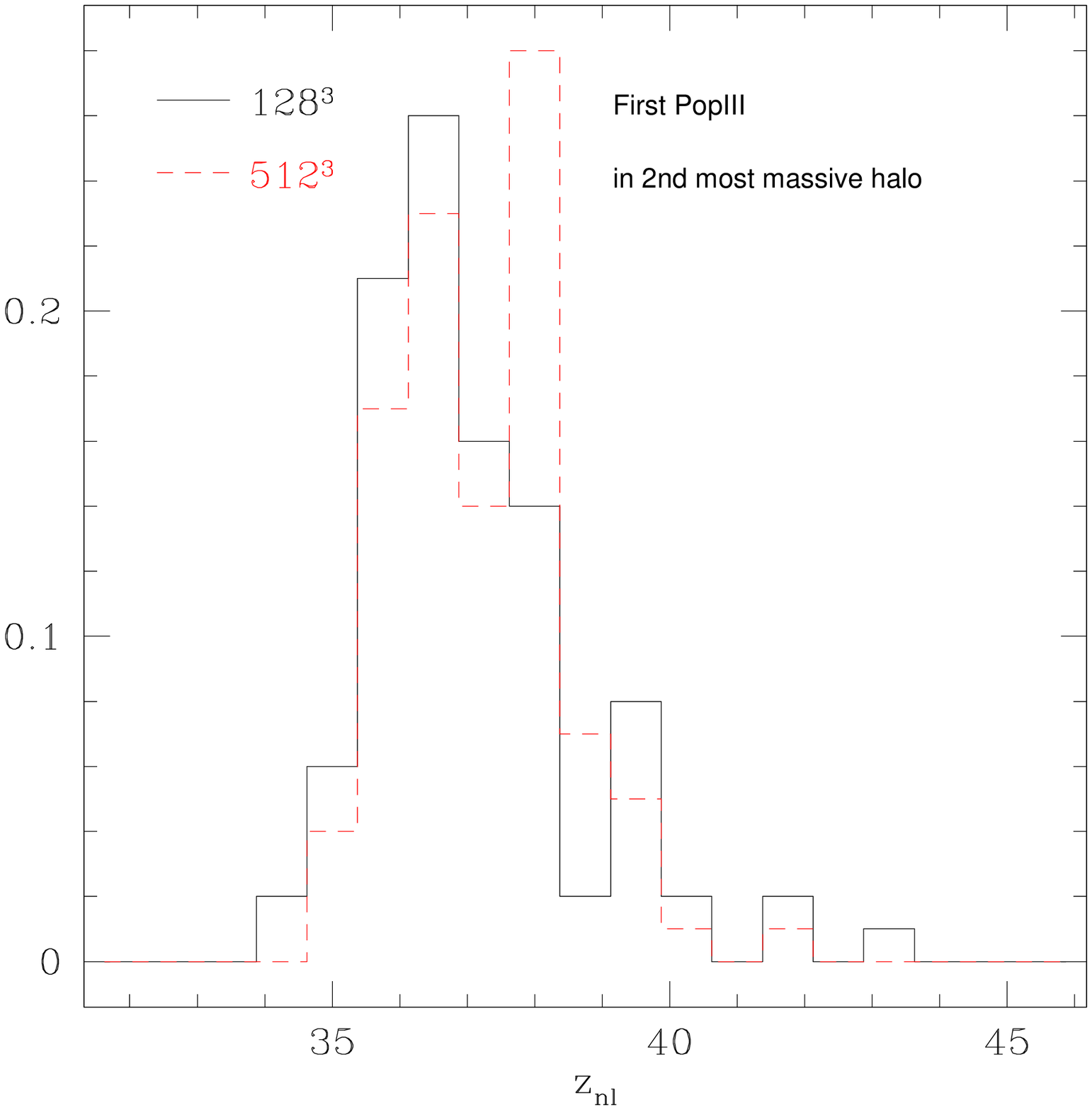} \caption{Probability distribution for
  the earliest PopIII star in the two most massive halos of the $S1$
  run obtained from the full resolution (red dashed line) and from a
  lower resolution constrained realization of the same initial
  conditions (solid black).}\label{fig:checkHALO}
\end{figure}

To verify the validity of our Monte Carlo method we perform two main
tests. First we compare the maximum overdensity at the first light
halo mass scale identified over the whole simulation box using
different grid resolutions, including the analytical expectation (that
is equivalent to assume that the whole box is just one cell). The
results are reported in Fig.~\ref{fig:check} and confirm indeed that
the method is independent of the grid size, with an excellent match
between all the probability distributions computed. The figure has
been obtained by first generating a Gaussian random field with $\sigma
= 0.1$ on a $N=64^3$ grid and adopting $\sigma_{fs} = 0.008$ and
$k=4$. Then we progressively bin grid cell values to obtain lower
resolution versions of the original field. The variance in the low
resolution grids scales as $(N/64^3)^{1/2}$ and the values for
$\sigma_{fs}$ and $k$ are correspondingly increased. As a second test,
shown in Fig.~\ref{fig:checkHALO}, we have realized a constrained low
resolution ($N=128^3$) version of the initial conditions for our $S1$
simulation and we have then carried out the run down to $z=6$. From
the snapshot at $z=6$ we identify the most massive halos in this
simulations, verifying that there is a good spatial and mass match
with the original $512^3$ run. The redshift distribution of the first
PopIII progenitor for the most massive halos has been computed using
our method and compared with that of the original run. The agreement
is very good, especially considering that the dynamics of the dark
matter halos has been followed at a resolution $64$ times lower.

\clearpage
  
\clearpage

\clearpage

\clearpage

\clearpage

\begin{table}
\begin{center}
\caption{Simulations Summary}
\label{tab:sim}
\begin{tabular}{cccccc}
\tableline\tableline
ID & $L_{box}$ {\small{h/Mpc}} &$\sigma_{8}$ & $k^{1/3}_{ref}$ & $M_{fs}$ {\small{h/$M_{\sun}$}} & $N_{MC}$ \\
\tableline
S1 & 60  & 0.9  & 5  &$8.6~10^5$& $100$ \\
S2 & 60  & 0.75 & 5  &$8.6~10^5$& $100$ \\
M1 & 512 & 0.9  & 40 &$1.0~10^6$& $200$ \\
L1 & 720 & 0.9  & 57 &$1.0~10^6$& $600$ \\

 \tableline
\end{tabular}
\tablecomments{Summary table with the details of the numerical
  simulations carried out in this paper. The first column gives the
  simulation ID ; $L_{box}$ (second column) is the box size, while the
  third column contains the value of $\sigma_8$ used to generate the
  initial conditions. $k_{ref}$ (fourth column) is the refinement
  factor from the mass of a single particle in the N-body run to the
  mass assumed for a PopIII star halo ($M_{fs}$; fifth column). The
  last column ($N_{MC}$) reports the number of different Monte Carlo
  realizations of the PopIII halo density field for each N-body
  realization.}
\end{center}
\end{table}


\end{document}